\documentclass[preprint,prd,noshowpacs]{revtex4}
\usepackage{amssymb}
\usepackage{amsmath}

\begin{document}

\title{A new look at the Dirac quantization condition
}

\author{Michael Dunia}
\email{michaelrobertdunia@mail.fresnostate.edu}
\affiliation{Physics Department, CSU Fresno, Fresno, CA 93740 USA}
\author{P.Q. Hung}
\email{pqh@virginia.edu}
\affiliation{Department of Physics, University of Virginia, Charlottesville, VA 22904-4714, USA}
\author{Douglas Singleton}
\email{dougs@csufresno.edu}
\affiliation{Physics Department, California State University Fresno, Fresno, CA 93740 USA\\}
\affiliation{Kavli Institute for Theoretical Physics, University of California Santa Barbara, Santa Barbara, CA 93106, USA}

\date{\today}

\begin{abstract}
The angular momentum of any quantum system should be {\it unambiguously} quantized. We show that such a quantization fails for a pure Dirac monopole due to a previously overlooked field angular momentum from the monopole-electric charge system coming from the magnetic field of the Dirac string and the electric field of the charge. Applying the point-splitting method to the monopole-charge system yields a total angular momentum which obeys the standard angular momentum algebra, but which is gauge {\it variant}. In contrast it is possible to properly quantize the angular momentum of a topological 't Hooft-Polyakov monopole plus charge. This implies that pure Dirac  monopoles are not viable -- only 't Hooft-Polyakov monopoles are theoretically consistent with angular momentum quantization and gauge invariance.   
\end{abstract}

\maketitle

\section{Dirac string}

Despite the lack of experimental support, the study of magnetic charge continues to be an area of active research. This is partly because magnetic charge gives an additional, dual symmetry to classical electrodynamics. Also having even one magnetic charge in the Universe would give an explanation of the quantization of electric charge via the Dirac condition $q g = \frac{n \hbar}{2}$ \cite{dirac,dirac1} where $q, g$ are the electric and magnetic charge respectively and $n$ is an integer. This continued interest can be seen in the founding of the MoEDAL collaboration, which has as one of its goals observing monopoles at the LHC \cite{moedal}. There is a review article from 2020 on magnetic charge \cite{mavromatos} which is a summary and update of all that is known about monopoles. Finally there is recent work \cite{hung1,hung2} which shows that electroweak, 't Hooft-Polyakov-like monopoles \cite{thooft,polyakov} could be embedded in the Standard Model, with masses that make observing these electroweak monopoles feasible at LHC energies.

A magnetic charge $g$ would naively have a Coulomb magnetic field ${\bf B} = \frac{g {\bf r}}{r^3}$ which implies $\nabla \cdot {\bf B} = 4 \pi g \delta ({\bf r})$. This last equation runs afoul of the relationship between the magnetic field and the vector potential namely ${\bf B} = \nabla \times {\bf A}$. For a well behaved ${\bf A}$ one has $\nabla \cdot {\bf B} = \nabla \cdot (\nabla \times {\bf A}) = 0$. 

The Dirac string potential \cite{dirac,dirac1} in spherical polar and cylindrical coordinates is
\begin{equation}
\label{A-coulomb}
    {\bf A}^{g} _\pm (r) = \frac{g}{r} \left( \frac{\pm 1 - \cos \theta }{\sin \theta} \right) {\bf{\hat \varphi}} ~ =\frac{g}{\rho} \left( \pm 1 - \frac{z}{\sqrt{\rho^2+z^2}} \right) {\bf{\hat \varphi}} ~.
\end{equation}
Naively taking the curl of \eqref{A-coulomb} gives a Coulomb magnetic field, $\nabla \times {\bf A}^{g} _\pm (r) = \frac{g {\bf r}}{r^3}$. The vector potentials in \eqref{A-coulomb} are related by the gauge transformation ${\bf A}^{g} _+ (r) = {\bf A}^{g} _- (r) + \nabla \alpha$ with a non-single valued gauge function $\alpha = 2 g \varphi$.

${\bf A}^{g} _\pm (r)$ is everywhere singular along the $\mp z$  axis. This string singularity in ${\bf A}^g _\pm$ can be regularized by defining the vector potential as \cite{heras}
\begin{equation}
\label{A-coulomb-string}
   {\bf A}^{regular}_{\pm} = \frac{g \Theta (\rho -\epsilon)}{\rho} \left( \pm 1 - \frac{z}{\sqrt{\rho^2+z^2 + \epsilon^2}} \right) {\bf{\hat \varphi}}~,
\end{equation}
where $\epsilon$ is a small quantity taken to zero at the end and $\Theta (x)$ is the standard step function: $\Theta (x) =1$ for $x \ge 1$ and $\Theta (x) =0$ for $x<0$.  
Taking the curl of ${\bf A}^{regular}_{\pm}$ and taking the limit $\epsilon \to 0$ gives (see details in appendix D of \cite{heras})
\begin{eqnarray}
\label{b-coulomb}
   {\bf B} =  \lim_{\epsilon \to 0} \nabla \times ({\bf A}^{regular} _\pm ) &=&  g \frac{{\bf \hat r}}{r^2} \pm 2 g \frac{\delta (\rho)}{\rho} \Theta (\mp z) {\bf {\hat z}} \nonumber \\
   &=& g \frac{{\bf \hat r}}{r^2} \pm 4 \pi g \delta (x) \delta (y) \Theta (\mp z) {\bf {\hat z}} ~,
\end{eqnarray}
where we have used $\frac{\delta(\rho)}{2 \pi \rho} = \delta (x) \delta (y).$
The first term on the right hand side of \eqref{b-coulomb} is the point, Coulomb part and the second term is the string contribution. Further discussion of this standard Dirac-string form of the monopole magnetic field given in \eqref{b-coulomb} can be found in various review articles \cite{heras,olive,blag}, monographs \cite{felsager,shnir} and research articles \cite{adorno}.
It is straightforward to check that using the magnetic field in \eqref{b-coulomb} yields $\nabla \cdot {\bf B} = 0$ {\it i.e.} the outward flux from the first, Coulomb term is balanced by the inward second, solenoid term.
The string contribution to ${\bf B}$ in \eqref{b-coulomb} ({\it i.e.} the $\pm 4 \pi g \delta (x) \delta (y) \Theta (\mp z) {\bf {\hat z}}$ term) can be obtained from a string vector potential of the form ${\bf A}^{string}_{\pm} = \pm \frac{2 g \Theta (\rho) \Theta (\mp z) }{\rho} {\bf{\hat \varphi}}$
\cite{heras,olive,blag,felsager,shnir,adorno}

At this point one imposes the Dirac veto -- the condition that a charged particle, in the presence of a monopole, does not ``see" the string part of the magnetic field in \eqref{b-coulomb}. There are many ways to do this as given in \cite{heras}, all of which lead to the Dirac quantization condition $qg= n \frac{\hbar}{2}$. Almost all of the methods for obtaining the Dirac condition require putting some condition on wavefunction of the charge $q$, {\it e.g.} that the wavefunction remain single valued as it circles around the Dirac string, or that the wavefunction vanish at the location of the string. However, there is an approach to obtaining the Dirac quantization condition which does not rely on the wavefunction of $q$, but rather uses the fact that the electric and magnetic fields of a charge-monopole system carry a field angular momentum. It is this approach to the Dirac quantization condition that we will focus on in this paper. We stress that this field angular momneutm apporach to the Dirac quantization condition may be considered more fundamental, since it depends on the quantization of angular momentum as opposed to placing an (arbitrary) condition on the wavefunction of $q$.  

\section{Total field angular momentum of the Dirac String}

 In this section we discuss the angular momentum approach to the Dirac quantization condition. We begin by placing the magnetic charge at the origin, so that the ${\bf B}$-field is given by \eqref{b-coulomb},  and we place the electric charge at the location ${\bf r}_0$  so that the electric field is ${\bf E} = q \frac{{\bf\hat r}'}{{r'}^2}$ (where ${\bf r}' = {\bf r} - {\bf r}_0$). Then the field angular momentum coming from the point part of the magnetic field ({\it i.e.} $g\frac{{\hat {\bf r}}}{r^2}$) is given by the well-known result \cite{heras,olive,blag,felsager,shnir,saha,saha1,wilson,fierz,lipkin} 
\begin{equation}
\label{ang3d}
{\bf L}^{point}_{EM}= \frac{1}{4 \pi} \int {\bf r} \times ({\bf E} \times {\bf B}) d^3 x = \frac{qg}{4 \pi} \int r {\bf\hat r} \times \left( \frac{{\bf r}'}{{r'}^3} \times {\frac{{\bf \hat r}}{r^2}} \right) d^3 x =  - qg {\hat {\bf r}_0} ~.
\end{equation} 
Taking the magnitude of \eqref{ang3d} and imposing the quantum mechanical requirement that all angular momentum must come in integer multiples of $\frac{\hbar}{2}$ one quickly obtains the Dirac condition, $qg = n \frac{\hbar}{2}$ \cite{saha,saha1,wilson}. This field angular momentum approach to the Dirac quantization condition uses the electric field of $q$. In contrast, other methods for obtaining the Dirac condition, such as those cataloged in \cite{heras} or the fiber bundle approach of Wu and Yang \cite{wu-yang}, rely on imposing conditions on the wavefunction of $q$.

However the string part of the magnetic field in \eqref{b-coulomb} ({\it i.e.} the $\pm 4 \pi g \delta (x) \delta (y) \Theta (\mp z) {\bf {\hat z}}$ term) also contributes to the field angular momentum. It is this contribution to the field angular momentum which is the focus of this work and which we claim has been overlooked in all previous work. It leads to new results for the Dirac string formulation of magnetic charge. This string contribution to the field angular momentum is 
\begin{eqnarray}
\label{ang3d-2a}
{\bf L}_{EM} ^{string} &=& \frac{1}{4 \pi} \int {\bf r} \times \left (q  \frac{{\bf \hat r}'}{{r'}^2} \times ( \pm 4 \pi g \delta (x) \delta (y) \Theta (\mp z) {\bf {\hat z}}) \right) d^3 x \nonumber \\
&=& \mp gq \left( \frac{z_0 \mp \sqrt{\rho_0^2+z_0^2}}{\rho_0} \right) {\bf{\hat \rho}_0} = \mp gq \left( \frac{\cos \theta _0 \mp 1}{\sin \theta_0} \right) {\bf{\hat \rho}_0}~,
\end{eqnarray} 
where ${\bf r}_0 = \rho_0 {\bf{\hat \rho}_0} + z_0 {\bf {\hat z}}$ is the location of the charge $q$. Unlike, ${\bf L}_{EM} ^{point}$, the string contribution ${\bf L}_{EM} ^{string}$ depends on the direction ({\it i.e.} $\pm z$) of the string, and thus is gauge variant since one has a different result depending on which of the two gauge related potential -- ${\bf A}^g _+$ or ${\bf A}^g _-$ -- one uses.  

Also, unlike, ${\bf L}_{EM} ^{point}$, the string contribution to the field angular momentum, ${\bf L}_{EM} ^{string}$, and therefore the total field angular momentum will depend on the angular coordinate, $\theta_0$. Converting \eqref{ang3d} to cylindrical coordinates ({\it i.e.} ${\bf L}^{point}_{EM} = - qg (\sin \theta_0 {\hat {\bf \rho} _0} + \cos \theta _0  {\hat {\bf z}})$) and combining this with \eqref{ang3d-2a} yields
\begin{eqnarray}
\label{L-total}
    {\bf L}^{total}_{EM} = {\bf L}_{EM} ^{point} + {\bf L}_{EM} ^{string}  &=&  - q g \cos \theta _0 {\bf {\hat z}} + qg (\mp 1 + \cos \theta_0 )\cot \theta_0 {\bf{\hat \rho}}_0 \nonumber \\
    &=& - q g \left(\frac{z_0}{r_0} \right) {\bf {\hat z}} + qg \left(\mp 1 + \frac{z_0}{r_0}  \right)\frac{z_0}{\rho^2_0} (x_0 {\bf \hat x}+ y_0 {\bf \hat y} )~,
\end{eqnarray}
In the second line of \eqref{L-total} we have written the result with Cartesian unit vectors and Cartesian coordinates for later use. 
The magnitude of this total field angular momentum using the first line of  \eqref{L-total} is $|{\bf L}^{total}_{EM}| = qg \cos(\theta _0) \sec (\theta_0 /2)$ for the $(-)$ case and $|{\bf L}^{total}_{EM}| = qg \cos(\theta _0) \csc (\theta_0 /2)$ for the ($+$) case. Thus the magnitude of ${\bf L}^{total}_{EM}$ varies with the position of $q$ via the $\theta_0$-dependent terms and one cannot use the heuristic angular momentum quantization approach \cite{saha,saha1,wilson} to deriving the Dirac quantization condition.

\section{Angular momentum commutators} 

One should be cautious about how seriously to take this breakdown of the field angular momentum method of references \cite{saha,saha1,wilson} for deriving the Dirac quantization condition. One should instead see if the total angular momentum -- particle plus field -- satisfies the proper commutation relationship for angular momentum namely, $[J^i , J^j] = i \hbar \epsilon ^{ijk} J^k$, which represent the rotational symmetry of the system. Such an analysis was carried out in \cite{fierz,lipkin,yang} as well as reviewed in \cite{olive,blag,heras} and it was found that when one combined the particle angular momentum with ${\bf L}^{point} _{EM}$ of \eqref{ang3d} that indeed ${\bf J} = ({\bf x} \times {\bf \Pi}) + {\bf L}^{point} _{EM}$ did satisfy $[J^i , J^j] = i \hbar \epsilon ^{ijk} J^k$. (Here ${\bf \Pi} = {\bf p} - q {\bf A}$ with ${\bf p}$, being the momentum operator and ${\bf A}$ the vector potential). We now check if the added string field angular momentum from \eqref{ang3d-2a} leads to a total angular momentum that is a good quantum angular momentum in that it satisfies $[J^i , J^j] = i \hbar \epsilon ^{ijk} J^k$. The new, total angular momentum is ${\bf J} = ({\bf x} \times {\bf \Pi}) + {\bf L}^{point} _{EM} + {\bf L}^{string} _{EM}$. To carry out the computations we write out each term in the total angular momentum in Cartesian coordinates with index notation. The first term in ${\bf J}$ is simply written as $\epsilon ^{ijk} x^j \Pi^k$. Similarly the usual field angular momentum term for the charge-monopole system can be written as $ (L^{point} _{EM})^i = g \frac{x^i}{r}$ with $r=\sqrt{x^2+y^2+z^2}$. Without confusion, here and in the following we drop the $0$ subscript of \eqref{ang3d} and \eqref{ang3d-2a} which indicates the location of the charge $q$. The string field angular momentum in Cartesian coordinates is ${\bf L}^{string} _{EM} = \mp qg \left( \frac{z\mp r}{\rho^2} \right)(x {\bf {\hat x}} + y {\bf {\hat y}})$ with $\rho = \sqrt {x^2 +y^2}$. In index notation this becomes $ (L^{string} _{EM})^i  =\mp qg \left( \frac{x^3 \mp r}{\rho ^2} \right) (x^1 \delta ^{i1} + x^2 \delta ^{i2})$. Putting this all together gives the total angular momentum operator as $J^i = \epsilon ^{ijk} x^j \Pi^k - q g \frac{x^i}{r}  \mp qg \left( \frac{z \mp r}{\rho ^2} \right) (x \delta ^{i1} + y \delta ^{i2})$. In terms of the explicit components, one has for $J^x, J^y$ and $J^z$
\begin{equation}
\label{j-total-x}
J^x = ({\bf x} \times {\bf \Pi})_x - qg \frac{x}{r} \mp gqx\bigg(\frac{z\mp r}{\rho ^2}\bigg)
\end{equation}
\begin{equation}
\label{j-total-y}
J^y = ({\bf x} \times {\bf \Pi})_y - qg \frac{y}{r} \mp gqy\bigg(\frac{z\mp r}{\rho ^2}\bigg)
\end{equation}
\begin{equation}
\label{j-total-z}
J^z =  ({\bf x} \times {\bf \Pi})_z - qg \frac{z}{r}
\end{equation}
Using \eqref{j-total-x} \eqref{j-total-y} \eqref{j-total-z}  and some common commutator results like $[x^i, \Pi ^j] = i \delta ^{ij}$ , $[f (x^i), \Pi ^j] = i \partial ^j f(x^i)$, $[\Pi ^i, \Pi ^j] = i q \epsilon ^{ijk} B^k$ one can obtain the commutators $[J^i, J^j]$ for the $J^i$'s in \eqref{j-total-x} \eqref{j-total-y} \eqref{j-total-z} with the final result being 
\begin{equation}
\label{jyjz}
[J^y, J^z] =  i J^x
\end{equation}
\begin{equation}
\label{jzjx}
[J^z, J^x] =  i J^y ~,
\end{equation}
\begin{equation} 
\label{jxjy}
[J^x, J^y] = i J^z \pm  i qg  \pm iqz({\bf x}\cdot {\bf B}^{string}_{\pm})  ~,
\end{equation}
The right hand sides of equations \eqref{jyjz} and \eqref{jzjx} do match the $J_x$ and $J_y$ total angular momenta from \eqref{j-total-x} and \eqref{j-total-y} {\it i.e.} the angular momentum commutators are satisfied for  $[J^z, J^x]$ and $[J^y, J^z]$ for the total angular momentum defined in \eqref{j-total-y} and \eqref{j-total-x}. So far so good. However the last two terms in $[J^x, J^y]$ ({\it i.e.} $\pm iqz({\bf x}\cdot {\bf B}^{string}_{\pm})\pm igq$) spoils this commutator, which according to \eqref{j-total-z}  should only be $i ({\bf x} \times {\bf \Pi})_z - iqg \frac{z}{r}$. The existence of the anomalous term $\pm iqz({\bf x}\cdot {\bf B}^{string}_{\pm})$ is well known and is related to the presence of singular operator products. In quantum field theory Schwinger showed how to deal with such singular operator products via the point splitting method \cite{schwinger}. Zumino applied this point splitting technique \cite{zumino} to the quantum system of a charge and monopole to deal with the anomalies in the angular momentum commutator in \eqref{jxjy}. Details of these calculations can be found in \cite{blag,shnir}. The point-split version of the total angular momentum operators are (the subscript indicates point split operators)
\begin{eqnarray}
\label{j-point-split-1}
    J^i _{(PS)}=\lim_{\varepsilon\to 0}\epsilon^{ijk}x^j\frac{\varepsilon^k}{i\varepsilon^2}\left[1 - \exp \left(-i\frac{\textbf{p}\cdot{\varepsilon}}{2} \right) \exp \left( iq\int_{\textbf{x}-{\varepsilon}/2}^{\textbf{x}+\varepsilon/2}\textbf{A}\cdot d{\xi} \right) \exp\left(-i\frac{\textbf{p}\cdot{\varepsilon}}{2} \right) \right]~.
\end{eqnarray}
Expanding the exponential, taking the limit and using $\langle \epsilon^{ijk}\frac{\varepsilon^k\varepsilon^l}{\varepsilon^2}\rangle = \epsilon^{ijk}\delta^{lk}$ and $\Pi^k=p^k-qA^k$, we find
\begin{eqnarray}
\label{j-point-split-2}
    J^i_{(PS)}&=&\lim_{\varepsilon\to 0}\epsilon^{ijk}x^j\bigg((p^k-qA^k)-(qx^m\partial^mA^k)\bigg)\nonumber\\&=&\epsilon^{ijk}x^j\Pi^k-q\epsilon^{ijk}x^jx^m\partial^mA^k
\end{eqnarray}
Letting $q\epsilon^{ijk}x^jx^m\partial^mA^k=q\epsilon^{ijk}x^jx^m(\partial^mA^k-\partial^kA^m+\partial^kA^m)=q\epsilon^{ijk}x^jx^m(\epsilon^{mkp}B^p+\partial^kA^m)$, we obtain
\begin{eqnarray}
\label{j-point-split-3}
    J^i_{(PS)}=\epsilon^{ijk}x^j\Pi^k-q\epsilon^{ijk}x^jx^m\epsilon^{mkp}B^p-q\epsilon^{ijk}x^jx^m\partial^kA^m
\end{eqnarray}
The middle term in \eqref{j-point-split-3} can be shown to vanish 
giving finally
\begin{eqnarray}
\label{j-point-split-4}
    J^i _{(PS)}=\epsilon^{ijk}x^j\Pi^k-q\epsilon^{ijk}x^jx^m\partial^kA^m ~.
\end{eqnarray}
We now work out explicitly the components of $-q\epsilon^{ijk}x^jx^m\partial^kA^m$. 
The $x$-component is 
\begin{equation}
\label{J-extra-x}
   -q\epsilon^{1jk}x^jx^m\partial^kA^m = -qg \frac{zx}{\rho^2}\left(\pm 1 - \frac{z}{r}\right)  = -qg \frac{x}{r} \mp qgx \left( \frac{z \mp r}{\rho ^2}\right)
\end{equation}
The $y$-component is 
\begin{equation}
\label{J-extra-y}
   -q\epsilon^{2jk}x^jx^m\partial^kA^m = -qg \frac{zy}{\rho^2}\left(\pm 1 - \frac{z}{r}\right) = -qg \frac{y}{r} \mp qgy \left( \frac{z \mp r}{\rho^2}\right)~.
\end{equation}
Finally, the $z$-component is 
\begin{equation}
\label{J-extra-z}
   -q\epsilon^{3jk}x^jx^m\partial^kA^m =  -qg\frac{z}{r} \pm qg 
\end{equation}
Explicitly from \eqref{j-point-split-4} \eqref{J-extra-x} \eqref{J-extra-y} \eqref{J-extra-z}  we find that 
\begin{equation}
\label{J-Ps}
J^x _{(PS)} = J^x ~~~;~~~
J^y _{(PS)} = J^y ~~~;~~~
J^z _{(PS)} = J^z \pm qg, 
\end{equation}
Thus the point split angular momentum satisfies the angular momentum algebra
\begin{equation}
\label{PS-ang-algebra}
[J^i _{(PS)}, J^j_{(PS)}] = i \epsilon ^{ijk} J^k _{(PS)} ~.
\end{equation}
Equation \eqref{PS-ang-algebra} is the standard angular momentum commutation relationship for the point split angular momentum. To fully check the rotational symmetry of the charge-monopole system one would need to check that the angular momentum commutes with the Hamiltonian {\it i.e.} $[J^i, H]=0$ where the Hamiltonian is $H = \frac{1}{2m}(-i \nabla -e {\bf A})^2$ \cite{olive,blag,zumino}. Note this Hamiltonian is for the particle of charge, $e$, and mass, $m$, only. It does not contain the contribution of the pure ${\bf E}$ and ${\bf B}$ fields. If one takes the standard angular momentum operator and standard Hamiltonian operator one finds that $J^i$ and $H$ do not commute \cite{zumino}. However, as shown in \cite{blag,zumino} one can define a point split version of the Hamiltonian that, along with the point split version of the angular momentum, do commute {\it i.e.} $[J^i _{(PS)}, H_{(PS)}]=0$. The point split version of the angular momentum operators in \eqref{J-Ps} only differ from the standard angular momentum operators by the added constant term  $\pm qg$, which comes form the string contribution. Thus the point split Hamiltonian defined in \cite{blag,zumino} should still commute with the point split angular momenta in \eqref{J-Ps}. 

Since the point split versions of the angular momentum and Hamiltonian do commute one can find common eigenfunctions for these operators. For the monopole-charge system, but without taking into account the field angular momentum of the string, these monopole spherical harmonics were given in \cite{wu-yang-2} (see also \cite{yang} for a simple discussion of the monopole spherical harmonics). It is not clear if the monopole spherical harmonics of \cite{wu-yang-2} will also serve as eigenfunctions for the angular momenta which include the string angular momentum. We leave the study of the eigenfunctions of $J^i _{(PS)}$ and $H_{(PS)}$ for future work. 

For $J^z _{(PS)}$ in \eqref{J-extra-z} we find that the anomalous $\pm iqz({\bf x}\cdot {\bf B}^{string}_{\pm})$ is gone as was the case from previous works \cite{blag,shnir,zumino}. Next the point split $z$-component of the angular momentum contains the term $-qg \frac{z}{r}$, which is the usual field angular momentum coming from the magnetic Coulomb field of the monopole. 

However although ${\bf J}_{(PS)}$ now satisfies the angular momentum algebra, it still contains gauge {\it variant terms} {\it e.g.} the $\pm qg $ term in $J^z_{(PS)}$ or the $\mp qg \frac{xz}{\rho^2}$ term in $J^x _{(PS)}$. This gauge variance of ${\bf J}_{(PS)}$  calls into question {\it the ability to properly quantize} ${\bf J}_{(PS)}$ as one would expect from quantum mechanics. 

The above application of the point splitting method is reminiscent of the chiral anomaly of massless spinor electrodynamics as discussed in \cite{treiman}. For spinor electrodynamics one can define an axial current ${\cal J}^5 _\mu = {\bar \psi} \gamma_\mu \gamma _5 \psi$ which has a divergence of $\partial^\mu {\cal J}^5 _\mu = 2 i m {\bar \psi} \gamma_5 \psi$, where $m$ is the mass of the fermion field $\psi$. In the limit when the fermion mass goes to zero this axial current should be conserved, $\lim _{m \to 0} ~ \partial^\mu {\cal J}^5 _\mu = 0$. However certain one-loop triangle graphs spoil this conservation. In \cite{treiman} the point splitting method is used to construct a new axial current ${\tilde {\cal J}}^5 _\mu = {\cal J}^5 _\mu - \frac{q^2}{2 \pi} {\cal F}_{\mu \nu } A^\nu$, where ${\cal F}_{\mu \nu} =\frac{1}{2}\epsilon_{\mu \nu \rho \sigma} F^{\rho \sigma}$ is the dual field strength tensor of the Maxwell field strength tensor $F^{\rho \sigma}$. This new axial current, ${\tilde {\cal J}}^5 _\mu$, is conserved ($\partial ^\mu {\tilde {\cal J}}^5 _\mu=0$), but is not gauge invariant due to the presence of the vector potential in the term $\frac{q^2}{2 \pi} {\cal F}_{\mu \nu } A^\nu$. Thus one had ${\cal J}^5 _\mu$ which was gauge invariant but not conserved, while ${\tilde {\cal J}}^5 _\mu$ was conserved, but not gauge invariant. This is similar to the present case where the $\pm qg$ term in $J^z _{(PS)}$ restores the angular momentum algebra, but calls into question the gauge invariance of the string, since one arrives at different field angular momentum depending on which gauge equivalent vector potential, $A_+$ or $A_-$, one uses to calculate the string field angular momentum in \eqref{ang3d-2a}.  

\section{Dirac versus 't Hooft-Polyakov monopoles}

The above analysis of the Dirac monopole-electric charge system emphasizes the importance of the total angular momentum of the system over making the string singularity ``invisible" to the wavefunction of the charge, $q$. In some sense the angular momentum approach is more fundamental since it is connected with rotational symmetry rather than coming from some condition one imposes on the wavefunction of the charge. If one takes the angular momentum approach as the more fundamental way of arriving at the Dirac quantization condition from the monopole-charge system, then one finds the dueling problems of the non-closure of the angular momentum algebra versus the gauge invariance of the total angular momentum. This problem can be resolved by considering only 't Hooft-Polyakov-type configurations as the correct model of monopoles, since they have no string singularity and thus avoid the need to impose a condition on the wavefunction of $q$. Monopoles of the 't Hooft-Polyakov-type still have a field angular momentum that takes the form
${\bf J} _{tHP} = {\bf r} \times {\bf \Pi} + qg {\bf \hat r}$ {\it i.e.} particle plus point field contributions {\it but} no string field contribution since the 't Hooft-Polyakov monopole does not have a string singularity. ${\bf J} _{tHP}$ {\it satisfies} the angular momentum algebra and further one can {\it properly quantize} ${\bf J} _{tHP} \cdot {\bf \hat r} = qg = n \frac{\hbar}{2}$ leading to the Dirac condition without reference to the wavefunction.  

For the 't Hooft-Polyakov topological monopoles one can see that the angular momentum approach to obtaining the Dirac quantization condition takes precedence over the other approaches. For the canonical set up of the fields for the 't Hooft-Polyakov monopole the non-Abelian gauge fields are taken to have the Wu-Yang form, $A_i ^a \propto \epsilon _{iab} \frac{x^b}{r^2}$, and the scalar field takes the ``hedgehog" form, $\phi^a \propto \frac{x^a}{r}$ (note the mixing of Lorentz, indices {\it i,j} and group indices {\it a, b}). In this format there is no singularity in either the gauge or scalar fields so the quantization condition can only come from the quantization of angular momentum which in turn in tightly connected with the non-Abelian group structure. Now one can transform the 't Hooft-Polyakov monopole into a Dirac form via a {\it singular} gauge transformation. The details of this can be found in section 5.1.6 of reference \cite{shnir}. This singular gauge transformation is singular exactly along the $\pm z$-axis, and it transforms the Wu-Yang form of the non-Abelian gauge field into a Dirac-like form as in \eqref{A-coulomb}, and the scalar field is transformed from the hedgehog form to a $\phi ^a \propto \delta ^{a3}$ {\it i.e.} a constant field pointing in some particular direction in group space (taken to be the $3$-direction here). This new Dirac-like form for the 't Hooft-Polyakov monopole is said to be in the Abelian gauge, but here one can see that the string singularity in the gauge field really is a gauge artifact since one can always do the reverse gauge transform back to the hedgehog plus Wu-Yang form for the scalar and gauge fields. Further, since in the original configuration of hedgehog plus Wu-Yang form the quantization condition came from angular momentum quantization, when one transformed to the Abelian gauge for the scalar and gauge fields the wavefunction of the charge already does not interact with the string singularity. For the pure Dirac monopole one can move the string singularity around via a gauge transformation, but one cannot completely transform away the singularity as is the case for the 't Hooft-Polyakov monopole where one can go between the hedgehog form, where there is no singularity, and Abelian form, which has a singularity. Then for the pure Dirac monopole the string field angular momentum proves problematic since one either cannot close the angular momentum algebra, or one is left with a gauge {\it variant} result.      

From the above analysis the takeaway point is that 't Hooft-Polyakov monopoles are the only theoretically consistent monopoles. This should guide future searches to focus on looking for 't Hooft-Polyakov monopoles instead of point-like Dirac monopoles via processes such as those proposed in \cite{nature-22}.

\section{Summary and conclusions}

In this paper we studied an overlooked contribution to the field angular momentum of the charge-monopole system that comes from the interaction of the electric field of $q$ with the magnetic field of the string given in \eqref{ang3d-2a}. This gives a total field angular momentum that is the sum of ${\bf L}^{point}_{EM}$ and ${\bf L}^{string}_{EM}$. This new string field angular momentum contribution spoils the simple argument for arriving at the Dirac quantization condition by requiring that the field angular momentum of the charge-monopole system take values of $n \frac{\hbar}{2}$ \cite{saha,saha1,wilson}. 
With the exception of the field angular momentum quantization method of \cite{saha,saha1,wilson} all other arguments leading to the Dirac quantization condition rely on the wavefunction of $q$.

We then turned to a more rigorous investigation of this new field angular momentum by calculating the total angular momentum commutators consisting of the canonical ${\bf x} \times {\bf \Pi}$ term plus the field angular momentum coming from the point and string contributions. These commutators had anomalous terms which spoiled the closure of the angular momentum commutators. However, these anomalies, and their remedy via defining a point split version of the angular momentum, were already well known via the works \cite{blag,shnir,zumino}. We found that the point split version of the total angular momentum in \eqref{j-point-split-1} and \eqref{j-point-split-4} did satisfy  the standard angular momentum algebra. However, the point-split angular momentum was found to be gauge {\it variant}. The demonstration that the point split angular momenta $J^i _{(PS)}$ obeyed the correct angular momentum algebra ({\it i.e.} \eqref{PS-ang-algebra}) and that the point split Hamiltonian commutated with the point split angular momentum ({\it i.e.}  $[J^i _{(PS)}, H_{(PS)}]=0$) did not take into account the quantum nature of the electromagnetic field. However, reference \cite{blag} shows that these results also carry over to quantum fields. 

The above calculations, which take into account the previously overlooked field angular momentum of the Dirac string, cast doubt on the consistency of the pure Dirac monopole along the lines of the end note 11 in reference \cite{zumino}. The present analysis shows that pure Dirac monopoles either violate the angular momentum algebra or lead to a gauge variant total angular momentum. This implies that the proper context for magnetic charges is as 't Hooft-Polyakov-type configurations \cite{polyakov}, where the string singularity of the pure Dirac monopole is replaced by a non-singular configuration of gauge plus scalar fields with a non-trivial topology. The Dirac quantization condition outside this topological monopole  comes directly from the unambiguous quantization of the total angular momentum of the monopole-charge system. As a final word of caution in the original 't Hooft-Polyakov monopole solution one has an unrealistic, toy model for the electroweak interaction ({\it i.e.} the gauge group is $SO(3)$). In this case the topological quantization happens to coincide with the Dirac quantizaiton condition. This is not always the case -- see the work in references \cite{hung1,hung2} where the topological quantization {\it is not} the same as the standard Dirac quantization, which allows for the prediction of the weak mixing angle \cite{hung2}. \\

{\bf Acknowledgment:} DS is a 2023-2024 KITP Fellow at the Kavli Institute for Theoretical Physics. This research was supported in part by the National Science Foundation under Grant No. NSF PHY-1748958.

\end{document}